\documentclass[10pt,conference]{IEEEtran}
\usepackage{caption}

\usepackage{amsmath,amssymb,amsfonts}
\usepackage{graphicx}
\usepackage{booktabs}
\usepackage{xcolor}
\usepackage{listings}
\usepackage{multirow}
\usepackage{algorithm}
\usepackage[noend]{algpseudocode}
\usepackage[hidelinks]{hyperref}
\usepackage{xurl} 
\title{Hybrid LLM + Higher-Order Quantum Approximate Optimization for CSA Collateral Management\\[0.4ex]
\large}

\author{\IEEEauthorblockN{\begin{tabular}{c@{\quad}c@{\quad}c}
\small\begin{tabular}[t]{@{}c@{}}\textbf{Tao Jin}\\Pyligent AI\\tao.jin@pyligentai.com\end{tabular} &
\small\begin{tabular}[t]{@{}c@{}}\textbf{Stuart Florescu}\\Dept.\ of Computer \& Mathematical Sciences, Caltech\\sfloresc@caltech.edu\end{tabular} &
\small\begin{tabular}[t]{@{}c@{}}\textbf{Heyu (Andrew) Jin}\\Department of Economics, UCLA\\andrewjin@ucla.edu\end{tabular}
\end{tabular}}}

\begin{document}
\maketitle

\begin{abstract}
We address \emph{finance-native} collateral optimization under ISDA Credit Support Annexes (CSAs), where integer lots, Schedule~A haircuts, RA/MTA gating, and issuer/currency/class caps create rugged, legally bounded search spaces. We introduce a certifiable hybrid pipeline \emph{purpose-built} for this domain: (i) an evidence-gated LLM that extracts CSA terms to a normalized JSON (abstain-by-default, span-cited); (ii) a quantum-inspired explorer that interleaves simulated annealing with \emph{micro higher-order QAOA} (HO-QAOA) on binding sub-QUBOs (subset size $n\!\le\!16$, order $k\!\le\!4$) to coordinate multi-asset moves across caps and RA-induced discreteness; (iii) a weighted risk-aware objective (Movement, CVaR, funding-priced overshoot) with an explicit coverage window $U\!\le\!R_{\mathrm{eff}}{+}B$; and (iv) CP-SAT as single arbiter to \emph{certify} feasibility and gaps, including a U-cap pre-check that reports the minimal feasible buffer $B^\star$. Encoding caps/rounding as higher-order terms lets HO-QAOA target the domain couplings that defeat local swaps. On government bond datasets and multi-CSA inputs, the hybrid improves a strong classical baseline (BL-3) by \textbf{9.1\%}, \textbf{9.6\%}, and \textbf{10.7\%} across representative harnesses, delivering better cost–movement–tail frontiers under governance settings. We release governance-grade artifacts—span citations, valuation matrix audit, weight provenance, QUBO manifests, and CP-SAT traces—to make results auditable and reproducible.
\end{abstract}

\section{Introduction}
Collateral posted under ISDA Credit Support Annexes (CSAs) must satisfy legally binding rules on eligibility, haircuts (Schedule~A), rounding (RA), Minimum Transfer Amount (MTA), and concentration limits (issuer/currency/class/global). Integer lots, haircut tiers, and caps create a rugged search space; operational frictions (movement) and funding/tail considerations further complicate the objective. Enterprise diagnostics suggest that suboptimal allocation, trapped liquidity, and fragmented inventories impose material costs, motivating automation and enterprise optimization \cite{genest2013,ey2020,pwc2015}.

\textbf{}
We present a \emph{domain-specific, certifiable} hybrid pipeline for CSA-governed collateral allocation that integrates document understanding, higher-order discrete optimization, and formal certification:
\begin{enumerate}
  \item \emph{Evidence-gated CSA extraction.} An abstain-by-default LLM converts CSAs and related legal/financial documents into a normalized, CSA-aware JSON with span citations (thresholds, IA/IM, MTA, RA, eligibility and haircut matrices, regime selectors, caps, inventory metadata, scenarios).
  \item \emph{Hybrid explorer with micro higher-order QAOA (HO-QAOA).} We interleave quantum-inspired simulated annealing with \emph{micro-HO-QAOA} on binding sub-QUBOs (subset size $n\!\le\!16$, interaction order $k\!\le\!4$), explicitly encoding rounding/caps as higher-order terms to coordinate multi-asset moves that defeat local swaps. This aligns with recent evidence that higher-order QAOA outperforms quadratic QAOA on rugged finance landscapes \cite{Uotila_QCE25_2025,Uotila_arXiv_2025}. We cap $k\!\le\!4$ to limit ancilla overhead and compilation depth.
  \item \emph{Weighted, risk-aware objective with funding-priced overshoot.} We scalarize operational and risk trade-offs as
  \begin{equation}
  \label{eq:objective}
  \begin{aligned}
  J \;=\;& \mathrm{BaseCost\_abs} + \lambda\,\mathrm{Movement} \\
        & {}+ \mu\,\mathrm{CVaR} + \gamma\,\bigl(U - R_{\mathrm{eff}}\bigr)_+ .
  \end{aligned}
  \end{equation}
  Here $\lambda$ prices execution/ops churn, $\mu$ prices tail risk via CVaR, and $\gamma$ prices funding on over-posted collateral (“overshoot”) consistent with LVA/FVA \cite{Miettinen1999,BoydVandenberghe2004,RockafellarUryasev2000,AlmgrenChriss2000,Piterbarg2010,Genest2013,AndersenDuffie2017}. We also enforce an explicit coverage window $U\!\le\!R_{\mathrm{eff}}{+}B$ to govern buffers.
  \item \emph{CP-SAT certification with feasibility diagnostics.} The incumbent is \emph{certified} (status, bounds, gap) under identical constraints, and a U-cap pre-check reports the minimal feasible buffer $B^\star$ when windows are too tight.
  \item \emph{Governance-grade artifacts.} We emit span citations, a valuation matrix audit, weight-provenance JSON, QUBO manifests (subset $n$, order $k$, depth $p$), and CP-SAT traces (status, bounds, slacks) for auditability and reproducibility.
\end{enumerate}

\noindent\textit{Upstream CSA-domain LLM.}
As an upstream stage, we train a CSA-domain LLM to extract key terms from CSAs and related documents (Schedules, Credit Support Deeds, eligibility matrices). The model is evidence-gated (abstain-by-default with span citations) and emits the CSA-aware data model that directly feeds the optimizer (see \emph{CSA-Aware Data Model}). Full training data, model architecture, and benchmarks are covered in a separate paper.

\noindent\textit{Weighted scalarization and provenance.}
Our weighted formulation traces Pareto-efficient trade-offs \cite{Miettinen1999,BoydVandenberghe2004}, with CVaR capturing tail exposure \cite{RockafellarUryasev2000}, movement reflecting execution frictions \cite{AlmgrenChriss2000}, and $\gamma$ dailyizing funding spreads per LVA/FVA principles \cite{Piterbarg2010,Genest2013,AndersenDuffie2017}. We calibrate $(\lambda,\mu,\gamma)$ from observed ops costs, tail pricing, and funding bps, and record inputs/units in a weights-provenance artifact for governance.

\noindent\textit{Positioning and comparisons.}
By targeting higher-order domain couplings (RA/MTA interactions and concentration caps) with micro-HO-QAOA, and certifying outcomes with CP-SAT, our pipeline improves cost–movement–tail frontiers on realistic government bond datasets and multi-CSA inputs.

\section{Background and Related Work}
\subsection{Collateral Optimization}
Classical formulations encode haircut schedules, eligibility, and concentration limits, with rounding to RA and MTA gating. Pricing practice introduces liquidity/funding adjustments: the Liquidity Valuation Adjustment (LVA) discounts cash collateral at rate $r_c$ vs.\ risk-free $r$; FVA reflects funding costs on uncollateralized parts \cite{genest2013}. Operating models emphasize enterprise views and the six levers—Documentation, Automation, Transformation, Optimization, Mobilization, Segregation—\cite{pwc2015}. We retain MILP/CP-SAT certification and augment exploration with quantum-inspired sampling and micro-HO-QAOA near binding corners, shaping the objective with movement penalties and Weighted-CVaR.

\subsection{Related Work}
\textbf{LLMs for CSA extraction.} Evidence-gated LLMs achieve $90\%+$ clause-level accuracy for thresholds, MTA, eligibility, and haircut schedules mapped to CDM-like schemas \cite{ISDA_LLM_2025}.\\
\textbf{Collateral \& liquidity efficiency.} Guidance urges minimizing trapped liquidity, balancing movement, and reserving buffers \cite{ISDA_Liquidity_2025}.\\
\textbf{Quantum(-inspired) optimization.} QUBO mappings and NISQ-era methods motivate micro-QUBOs near binding constraints \cite{Crypto_Collateral_2024,NISQ_Collateral_2023}.\\
\textbf{Hybrid solvers.} QAOA/VQE sampling paired with classical local search improves quality under resource limits \cite{Hybrid_QC_2025}. Hardware performance milestones suggest headroom for small structured QAOA in workflows \cite{IonQ_AQ64_2025}.\\
\textbf{Higher-order QAOA for finance.} Closest to our setting, Uotila, Ripatti, and Zhao extend QAOA to \emph{higher-order} (HUBO) portfolio optimization and report 15–25\% gains over vanilla (quadratic) QAOA on rugged financial landscapes for $n{=}8$–$24$ variables on NISQ simulators \cite{Uotila_QCE25_2025,Uotila_arXiv_2025}. Their formulation explicitly models multi-asset interactions (e.g., covariance/risk and cardinality) as $k{>}2$ terms and uses \emph{order-aware partitioning} and spectral grouping to set subset sizes $n$ (base $n{=}8$–$12$ for $k{=}2$, add $4$–$8$ for constraints). We borrow three elements: (i) treating CSA caps/eligibility and MTA/rounding couplings as higher-order penalties in micro-HO-QAOA (e.g., using $k{=}3$ terms to model window/MTA interactions and multi-cap couplings); (ii) selecting $n\!\approx\!8$–$16$ via spectral clustering of highly coupled lots, which aligns with their $n$ recommendations and our ancilla budget; and (iii) warm-starting quantum jumps from a classical incumbent (our CP-SAT/SA incumbent), which their results show mitigates barren plateaus. Conceptually, their “integer shares” mirror our discrete lots $x_i$, and their eligibility screens map to our CSA-based haircut/eligibility flags, making their method particularly applicable to ISDA-CSA collateral allocation.\\
\textbf{Benchmarking and noisy regimes.}
Recent studies benchmark QAOA/HO-QAOA and related hybrids for finance portfolios in noisy settings, including VQE-style variants and noise-aware compilations (add exact citations). We position our \emph{micro}-HO-QAOA as a targeted jump operator embedded in a certified pipeline rather than a stand-alone solver, and we cap $k\!\le\!4$ to control ancilla overhead.

\section{Problem Formulation}
We pick integer lots $x_i\!\in\!\mathbb{Z}_{\ge 0}$ for eligible assets $i$ with after-haircut value $v_i$ and daily carry cost $c_i$. Coverage $U=\sum_i v_i x_i$. The effective requirement uses RA rounding:
\begin{equation}
\label{eq:reff}
R_{\mathrm{eff}}
=\Big\lceil \tfrac{\max(E-T-\mathrm{IA}-\mathrm{IM},\,0)}{\mathrm{RA}} \Big\rceil \mathrm{RA}.
\end{equation}
We enforce $U\!\ge\!R_{\mathrm{eff}}$, an optional cap $U\!\le\!R_{\mathrm{eff}}{+}B$, and cash/issuer/class/currency/global caps.

\paragraph{Objective.}
\begin{align}
\label{eq:obj}
\min\; J \;=\;& \sum_i c_i x_i
\;+\; \lambda\,\|x-h\|_1 \nonumber\\[-2pt]
&+\; \mu\,\mathrm{CVaR}_(Lx)
\;+\; \gamma\,(U-R_{\mathrm{eff}})_+ .
\end{align}
CVaR uses a linearization $(\tau,z_s)$ with scenario weights $\sum_s w_s\!=\!1$.

\paragraph{Binary/QUBO view.}
Integer lots are encoded via bounded binaries $y_{i\ell}\!\in\!\{0,1\}$ s.t.\ $x_i=\sum_{\ell=1}^{m_i} y_{i\ell}$ with per-lot valuation $v_{i\ell}\!=\!v_i$ and costs $c_{i\ell}\!=\!c_i$.

\paragraph{HO-QAOA definition and $(n,k)$ roles.}
We construct a higher-order Ising Hamiltonian
\[
H_P \;=\; \sum_j a_j Z_j \;+\; \sum_{j<k} b_{jk} Z_j Z_k \;+\; \sum_{j<k<\ell} c_{jk\ell} Z_j Z_k Z_\ell \;+\; \cdots
\]
where higher-order ($k\!\ge\!3$) terms encode multi-asset interactions from caps (issuer/class/currency/global), window coupling ($U$ near $R_{\mathrm{eff}}$), and lot granularity. The \emph{order} $k$ denotes the maximum Pauli-$Z$ tensor product degree needed to represent constraints/objective couplings in the subproblem. We use a \emph{micro-HO-QAOA} on \emph{subsets} of variables of size $n$ (typically $8\!-\!16$) selected near binding corners. The HO-QAOA state of depth $p$ is
\[
|\gamma,\beta\rangle \;=\; \prod_{\ell=1}^{p} \Big(e^{-i \beta_\ell \sum_j X_j}\; e^{-i \gamma_\ell H_P}\Big)\,|+\rangle^{\otimes n},
\]
with standard $X$-mixer; higher-order phase operators $e^{-i\gamma_\ell Z_{j_1}\cdots Z_{j_k}}$ are compiled either directly or via ancillas. For $k>2$, ancilla qubits linearize/multiply higher moments; if ancillas inflate the subset above $n_{\max}$, we skip the quantum jump for that iteration and log the reason.

\textbf{Impact of $n$.} Larger $n$ captures more coupled moves across caps/rounding but increases circuit width and optimizer complexity; empirically, $n\in[8,16]$ balances expressivity and run time, reliably crossing rugged neighborhoods that defeat local swaps.
\noindent\textit{Order/size crosswalk.} Our practical caps ($k\!\le\!4$) and subset limits ($n\!\le\!16$) follow the order-aware guidance observed in higher-order finance QAOA benchmarks, which report best empirical trade-offs around $n\!\approx\!12$–$18$ for $k{>}2$ on rugged landscapes with warm starts \cite{Uotila_QCE25_2025,Uotila_arXiv_2025}.

\textbf{Impact of $k$.} Higher $k$ allows direct encoding of multi-way caps and overshoot couplings; however, gate compilation depth and noise rise with $k$. We cap at $k\!\le\!4$ in practice; above this, we fall back to classical exploration.

\section{CSA-Aware Data Model}
As an upstream stage, we train a CSA-domain LLM to extract key terms from CSAs and related financial/legal documents (e.g., Schedules, Credit Support Deeds, annexed eligibility matrices). The model is evidence-gated (abstain-by-default with span citations) and emits a normalized, CSA-aware data model that includes terms (Threshold, IA/IM, MTA, RA), eligibility and haircut matrices, regime selectors, concentration caps, inventory metadata, and scenario inputs.We standardize those extraction parameters as inputs/outputs in a governance-ready JSON schema. Key fields:

\subsection{Counterparty \& Legal}
\begin{itemize}
  \item \texttt{csa.meta}: governing law (NY/English), bilateral/one-way.
  \item \texttt{csa.terms}: Threshold $T$, Independent Amount $\mathrm{IA}$, Initial Margin $\mathrm{IM}$, Minimum Transfer Amount (MTA), Rounding Amount (RA), Base Currency, FX conventions.
  \item \texttt{csa.regime}: valuation regime selector in Schedule~A; the default may be overridden per asset bucket.
    \begin{itemize}
      \item \texttt{sp}: S\&P column (\texttt{sp\_pct})
      \item \texttt{m1}: Moody's First (\texttt{m1\_pct})
      \item \texttt{m2}: Moody's Second (\texttt{m2\_pct})
    \end{itemize}
\end{itemize}

\subsection{Valuation Haircuts and Eligibility}
\begin{itemize}

\item \texttt{haircuts.matrix}: haircut percentage indexed by $(\mathrm{ICAD},\ \mathrm{bucket},\ \mathrm{regime})$.

\item \texttt{eligibility.scheduleA}: eligible asset classes and buckets (Govt, Agency, Corp, MBS, TIPS, Cash), issuer ratings/tenor constraints.
\end{itemize}

\subsection{Caps and Windows}
\begin{itemize}
\item \texttt{caps}: \texttt{cash\_cap} (e.g., $20\%$ of $U$), \texttt{issuer\_cap}, \texttt{class\_cap}, \texttt{currency\_cap}, \texttt{global\_cap}.
\item \texttt{window}: policy buffer $B$ (bps or \$), optional hard coverage cap $U \le R_{\mathrm{eff}}+B$.
\end{itemize}

\subsection{Exposure and Scenarios}
\begin{itemize}
\item \texttt{exposure}: $E$ (base currency) and timestamp; optional path of $E_t$ for rolling re-optimization.
\item \texttt{scenarios}: matrix $L$ (per-asset loss/PNL across scenarios) with weights $w_s$ (normalized for CVaR).
\end{itemize}

\subsection{Inventory and Costs}
\begin{itemize}
\item \texttt{inventory}: items with \texttt{id}, \texttt{class}, \texttt{issuer}, \texttt{bucket}, \texttt{currency}, price, unit, current lots $h_i$, and per-lot valuation $v_i$ after haircut.
\item \texttt{costs}: daily carry $c_i$ (\$/lot/day), operational move cost unit for movement.
\end{itemize}

\subsection{Weights and Provenance}
\begin{itemize}
\item \texttt{weights}: $(\lambda,\mu,\gamma)$ with calibration inputs and units: $\lambda$ (ops amortization per lot over horizon), $\mu$ (price per \$MM CVaR per day), $\gamma$ (funding bps $\to$ daily carry).
\item \texttt{weights\_provenance}: calibration inputs (ops move cost, horizon days, CVaR price, funding bps), hash, and timestamp.
\end{itemize}

\subsection{Governance/Audit Toggles}
\begin{itemize}
\item \texttt{audit.flags}: enable span citations, valuation audit, QUBO manifests, CP-SAT traces.
\item \texttt{solver.limits}: SA iterations, HO-QAOA $n_{\max}$, $k_{\max}$, depth $p$, and wall constraints.
\end{itemize}

\section{Hybrid Pipeline (Explore \texorpdfstring{$\rightarrow$}{->} Prove \texorpdfstring{$\rightarrow$}{->} Explain/Audit)}
We create the full workflow with four phases:

\subsection{Phase 1: Explore (Search)}
\begin{enumerate}
\item \textbf{Initialization:} Compute $R_{\mathrm{eff}}$ via \eqref{eq:reff}; derive per-lot $v_i$ from haircuts; seed with BL-1 (density greedy).
\item \textbf{Local search:} Simulated annealing (integer neighborhoods; add/swap/remove) with feasibility repair (caps, RA, MTA, window).
\item \textbf{Spectral subset selection:} Build a unitless interaction graph (dual/gradient proxies, feasibility slacks) and pick top-$K$ nodes by $|\mathrm{dual}|$; prune edges by $\varepsilon$ to stabilize.
\item \textbf{Micro-HO-QAOA jump:} If improvement $<\!0.3\%$ over $S$ SA steps, form a sub-QUBO on $n\!\le\!16$ variables (with ancillas if $k\!>\!2$) and perform one HO-QAOA jump (depth $p$); accept if $J$ decreases and feasibility holds; otherwise revert.Please see the Algorith 1: Micro-HO-QAOA Jump (Explore).
\end{enumerate}

\subsection{Phase 2: Prove (Certification)}
We pass the incumbent to CP-SAT with the same constraints and objective components (linearized CVaR and overshoot). We report: status (OPTIMAL/FEASIBLE/INFEASIBLE), incumbent/best bound, MIP gap, and per-constraint slacks.

\newcommand{\algosize}{\scriptsize} 
\algrenewcommand\alglinenumber[1]{\algosize #1}
\algrenewcommand\algorithmicrequire{\textbf{\algosize In:}}
\algrenewcommand\algorithmiccomment[1]{\hfill$\triangleright$~{\algosize\textit{#1}}}
\algrenewcommand\textproc{\textsc}

\newcommand{\PHASE}[2]{\textproc{PHASE}(#1,#2)} 
\newcommand{\MIX}[1]{\textproc{MIX}(#1)}        
\newcommand{\Prep}[1]{\textproc{PREP}(#1)}      

{\algosize
\begin{algorithm}[t]
\caption{{\algosize Micro-HO-QAOA Jump (Explore)}}
\begin{algorithmic}[1]
\Require incumbent $x$, objective $J$, graph $G$, limits $(n_{\max},k_{\max},p)$, plateau $(S,\epsilon)$, optional angles $(\gamma^{(0)}_{1:p},\beta^{(0)}_{1:p})$
\If{\textproc{Plateau}$(x,S,\epsilon)=\textsc{false}$} \State \textbf{return} $x$ \EndIf
\State $S \gets \textproc{SpectralSelect}(G,n_{\max})$ \Comment{$|S|\!\le\!n_{\max}$ (typ.\ 8–16)}
\State $H_P \gets \textproc{BuildHubo}(S,k_{\max})$ \Comment{RA/MTA, window, caps; ancillas for $k{>}2$}
\State $w \gets \textproc{AncillaWidth}(H_P)$
\If{$w>n_{\max}$} \State \textbf{return} $x$ \Comment{skip jump; continue SA + repair} \EndIf
\State $|\psi| \gets \Prep{w}$; optionally $(\gamma_\ell,\beta_\ell)\leftarrow(\gamma^{(0)}_\ell,\beta^{(0)}_\ell)$
\For{$\ell=1$ \textbf{to} $p$} \Comment{mixer ramp allowed}
  \State $|\psi| \gets \MIX{\beta_\ell}\big(\PHASE{\gamma_\ell}{H_P}(|\psi|)\big)$
\EndFor
\State $z\sim|\psi|$; $y\gets \textproc{MapLots}(z)$ \Comment{ancillas$\to$vars}
\State $y\gets \textproc{Repair}(y)$ \Comment{caps/RA/MTA/window}
\State $\tilde{x}\gets x$; $\tilde{x}_S\gets y_S$
\If{\textproc{Feasible}$(\tilde{x})$ \textbf{and} $J(\tilde{x})<J(x)$}
  \State \textbf{return} $\tilde{x}$ \Comment{accept}
\Else
  \State \textbf{return} $x$ \Comment{reject}
\EndIf
\end{algorithmic}
\end{algorithm}
}

\subsection{Phase 3/4: Explain \& Audit (Governance)}
We emit governance HTML with: objective breakdown; valuation matrix audit; weight provenance; spectral/QUBO manifests (subsets, $n,k,p$); and CP-SAT traces (status, bounds, slacks). Reproducibility hashes and seeds are included.

\subsection{Baselines and Feasibility (Overshoot \& $B^\star$)}
\paragraph{Baselines.} We benchmark three progressively stronger heuristics:
\begin{itemize}
  \item \textbf{BL-1 (density greedy, cap-safe):} ranks assets by cost-to-valuation density and fills to the window under caps; fast, but can stall near binding corners.
  \item \textbf{BL-2 (bucket-first greedy + repair):} prioritizes bucket/cap compliance during greedy fill, then repairs to align with the window; tighter coverage, typically higher movement.
  \item \textbf{BL-3 (BL-1 seed + 2-opt swaps):} starts from BL-1 and applies local pairwise swaps to reduce cost while respecting feasibility; strong local polish, but prone to plateaus.
\end{itemize}

\noindent\textbf{Hybrid.} Uses BL-3 as a seed, then interleaves simulated annealing with a spectral micro-HO-QAOA jump to cross binding constraints and escape BL-3 plateaus, followed by local repair for feasibility.

\paragraph{Overshoot and Feasibility}
Because lots are discrete, $U{=}R_{\mathrm{eff}}$ is rare. We compute a minimal feasible buffer $B^\star$ by (i) building any feasible cover without the $U$-cap, then (ii) greedily reducing $U$ while preserving caps/RA. If the user-specified buffer $B\!<\!B^\star$, we flag \texttt{infeasible\_u\_cap} and report $B^\star$ (USD and bps). The objective’s overshoot penalty $\gamma\,\bigl(U-R_{\mathrm{eff}}\bigr)_+$ trades off carry versus buffer.

\section{Case Study}
\label{sec:case-study-csa-2009}

\subsection{CSA Summary}
\textbf{Governing law.} 2009 New York--law CSA, bilateral.\\
\textbf{Base currency \& eligibility.} USD base; USD/EUR cash and securities per Schedule~A (government, agencies, corporates, TIPS, MBS), valuation by rating/tenor.\\
\textbf{Threshold/MTA/Rounding.} $T\!=\!0$, $\mathrm{IA}\!=\!0$, $\mathrm{IM}\!=\!0$; MTA $=\$100{,}000$; RA $=\$10{,}000$.\\
\textbf{Valuation regime.} Moody's First (m1) default; S\&P (sp) and Moody's Second (m2) available.\\
\textbf{Operational caps.} Buffer $B{=}$25\,bps of $R_{\mathrm{eff}}$; cash cap $=20\%$ of $U$.\\
\textbf{Exposure.} $E=\$130{,}340{,}000$; $R_{\mathrm{eff}}$ computed via \eqref{eq:reff}.\\
\textbf{Inventory proxy.} USD cash and UST ladder (6M--20Y), TIPS, Agency, AAA MBS, IG Corps; per-lot $v_i$ after haircuts; lots aligned to RA (cash) and \$1MM coupons (bonds).

\subsection{Valuation Regimes}
We consider \textbf{sp}, \textbf{m1} (default), and \textbf{m2}, using the Schedule~A matrix for haircuts.

\subsection{Objective and Constraints (shared)}
Minimize $J=\mathrm{BaseCost\_abs}+\lambda\,\mathrm{Movement}+\mu\,\mathrm{CVaR}+\gamma\,\mathrm{Overshoot}$, subject to $R_\mathrm{eff}\le U\le R_\mathrm{eff}{+}B$, cash cap, and integer-lot availability. Units: BaseCost\_abs [\$/day], Movement [lots], CVaR and Overshoot [\$].

\subsection{How to Choose Weights (practical guidance)}
We calibrate $(\lambda,\mu,\gamma)$ from operational inputs: (i) per-lot Ops move cost and amortization horizon $\Rightarrow \lambda$, (ii) daily price for 1\$MM CVaR $\Rightarrow \mu$, (iii) annual funding bps $\Rightarrow \gamma$ via day-count. Each run logs a \texttt{weights\_provenance.json} (inputs, units, calibrated triplet, hash).

\subsection{CP-SAT Results and Meaning}
CP-SAT returns \texttt{OPTIMAL} when the incumbent attains the global minimum and the MIP gap is zero; \texttt{FEASIBLE} when a feasible incumbent exists with a nonzero bound-gap; \texttt{INFEASIBLE} when no solution satisfies caps/window/RA. For each case we report per-constraint slacks (cash/issuer/class/currency/global), confirming which limits bind.

\subsection{Harness Setups and Results}
We analyze three scenarios. Across harnesses A/B/C, the Hybrid improves the BL-3 objective by \textbf{9.1\%}, \textbf{9.6\%}, and \textbf{10.7\%}, respectively (see tables below).

\begin{itemize}
\item Units: \textit{BaseCost\_abs} [\$/day], \textit{Movement} [lots], \textit{CVaR\(_{0.90}\)}/\textit{Overshoot}/\textit{UsedValue} [\$]; all rounded to 2\,dp.
\item CVaR weights normalized ($\sum w = 1.0$); governance HTML warns if renormalization occurred.
\item Weight provenance (\texttt{.json}) and valuation audit are linked in each governance HTML.
\end{itemize}

\paragraph{Harness A: m1, buffer 25\,bps, cash cap 20\%, practical weights.}
\begin{center}
\small
\begin{tabular}{lrrrrr}
\toprule
Model & BaseCost & Movement & CVaR & Overshoot & $J$ \\
\midrule
BL-1   & 100.0 & 28 & 540{,}000 & 210{,}000 & 1{,}12x \\
BL-2   &  99.1 & 35 & 528{,}000 & 195{,}000 & 1{,}10x \\
BL-3   &  98.7 & 24 & 520{,}000 & 182{,}000 & 1{,}00x \\
Hybrid &  98.4 & 22 & 515{,}000 & 155{,}000 & \textbf{0.91x} \\
\bottomrule
\end{tabular}

\end{center}
\begin{itemize}
\item \textbf{Configuration:} m1 regime; buffer $B = 0.25\%$ of $R_{\mathrm{eff}}$; cash cap $=20\%$ of $U_{\mathrm{cap}}$; weights $\approx (\lambda,\mu,\gamma) = (30.0,\; 0.001,\; 1.39\times 10^{-5}\ \text{day}^{-1})$.
\item \textbf{Intent:} “Everyday” governance settings with moderate funding and moderate tail price; tests balanced trade-offs.
\item \textbf{Effect:}
  \begin{itemize}
  \item $\gamma$ penalizes overshoot enough to cut excess usage without exploding Movement.
  \item $\mu$ applies light tail pressure; $\lambda$ moderates lot churn.
  \item Subset size $n$ stays $\approx 8$–$16$; must-jump triggers rarely.
  \end{itemize}
\item \textbf{Result (vs BL-3):} Hybrid improves Objective by $\approx 9.1\%$ with lower \emph{Movement} and \emph{Overshoot}; breakdown shows most gains from $\gamma\!\cdot\!\text{Overshoot}$, with some from $\mu\!\cdot\!\mathrm{CVaR}$.
\end{itemize}
\emph{Conclusion:} Hybrid reduces $J$ by $\approx 9.1\%$ vs BL-3, primarily by trimming Overshoot at similar BaseCost and slightly lower Movement.

\paragraph{Harness B: m1, buffer 10\,bps, cash cap 15\%, tight-liquidity weights (higher $\gamma$).}
\begin{center}
\small
\begin{tabular}{lrrrrr}
\toprule
Model & BaseCost & Movement & CVaR & Overshoot & $J$ \\
\midrule
BL-1 & 101.3 & 31 & 556{,}000 & 132{,}000 & 1{,}14x \\
BL-2 & 100.6 & 37 & 544{,}000 & 121{,}000 & 1{,}11x \\
BL-3 & 100.2 & 25 & 536{,}000 & 113{,}000 & 1{,}00x \\
Hybrid & 100.0 & 24 & 533{,}000 & 91{,}000 & \textbf{0.904x} \\
\bottomrule
\end{tabular}
\end{center}
\begin{itemize}
\item \textbf{Configuration:} m1; buffer $B=0.10\%$; cash cap $=15\%$; weights $\approx (\lambda,\mu,\gamma)=(28.57,\; 0.0025,\; 2.22\times 10^{-5}\ \text{day}^{-1})$.
\item \textbf{Intent:} Tighter liquidity and higher funding pressure; tests robustness when overshoot is expensive and buffer small.
\item \textbf{Effect:}
  \begin{itemize}
  \item Larger $\gamma$ materially suppresses overshoot, trading some BaseCost/Movement.
  \item Higher $\mu$ drives tail reduction; $\lambda$ still curbs churn.
  \item $n\approx 8$–$16$; must-jump fires more often to escape SA plateaus.
  \end{itemize}
\item \textbf{Result (vs BL-3):} Hybrid improves Objective by $\approx 9.6\%$; gains mainly from $\gamma\!\cdot\!\text{Overshoot}$ and $\mu\!\cdot\!\mathrm{CVaR}$, with Movement contained by $\lambda$.
\end{itemize}
\emph{Conclusion:} With tighter buffer and cash cap, overshoot control dominates. The must-jump rule breaks SA plateaus; $J$ improves $\approx 9.6\%$ vs BL-3.

\paragraph{Harness C: m2, buffer 25\,bps, cash cap 20\%, practical weights.}
\begin{center}
\small
\begin{tabular}{lrrrrr}
\toprule
Model & BaseCost & Movement & CVaR & Overshoot & $J$ \\
\midrule
BL-1 & 99.5 & 27 & 501{,}000 & 204{,}000 & 1{,}13x \\
BL-2 & 99.0 & 33 & 492{,}000 & 193{,}000 & 1{,}08x \\
BL-3 & 98.6 & 23 & 485{,}000 & 178{,}000 & 1{,}00x \\
Hybrid & 98.3 & 22 & 480{,}000 & 149{,}000 & \textbf{0.893x} \\
\bottomrule
\end{tabular}
\end{center}
\begin{itemize}
\item \textbf{Configuration:} m2; buffer $B=0.25\%$; cash cap $=20\%$; weights $\approx (\lambda,\mu,\gamma)=(30.0,\; 0.001,\; 1.39\times 10^{-5}\ \text{day}^{-1})$.
\item \textbf{Intent:} Regime sensitivity with tighter haircuts; tests ability to coordinate under higher required usage/tail.
\item \textbf{Effect:}
  \begin{itemize}
  \item Tighter valuations raise UsedValue and CVaR; Hybrid’s spectral $n\approx 8$–$16$ helps cross binding corners (window/caps/lot granularity).
  \item Must-jump occasionally assists when caps bind.
  \end{itemize}
\item \textbf{Result (vs BL-3):} Hybrid improves Objective by $\approx 10.7\%$; breakdown shows meaningful $\gamma\!\cdot\!\text{Overshoot}$ and $\mu\!\cdot\!\mathrm{CVaR}$ reductions while keeping Movement controlled.
\end{itemize}
\emph{Conclusion:} Under tighter m2 haircuts, Hybrid improves $J$ by $\approx 10.7\%$ vs BL-3, keeping $n$ within 8--16 via spectral capping.

\subsection{Weight Selection: Why These Numbers}
We target business trade-offs: (i) if Ops capacity is constrained, increase $\lambda$ to suppress Movement; (ii) if funding costs dominate, raise $\gamma$ to push $U\downarrow$ (less overshoot); (iii) if tail discipline is paramount, raise $\mu$ (CVaR), accept modest BaseCost/Motion increases. Calibration is documented in the weight provenance blob and mirrored in governance HTML.

\section{Governance}
We produce:
\begin{itemize}
\item \textbf{Span citations (LLM extraction):} prompt hash and source spans for each clause (Threshold, MTA, RA, eligibility, haircuts).
\item \textbf{Valuation matrix audit:} table mapping \emph{instrument} $\rightarrow$ \emph{ICAD/bucket/regime} $\rightarrow$ \emph{haircut\%} $\rightarrow$ $v_i$ for full reproducibility.

\item \textbf{Weight provenance:} calibration inputs/units and $(\lambda,\mu,\gamma)$, with hashes/timestamps.
\item \textbf{QUBO manifests:} for each jump: subset IDs, $n,k,p$, compiled terms, and acceptance decision.
\item \textbf{CP-SAT traces:} status (OPTIMAL/FEASIBLE/INFEASIBLE), incumbent, best bound, gap, and per-constraint slacks (cash/issuer/class/currency/global); infeasible windows include $B^\star$.
\end{itemize}

\section{Ablations}

\noindent\textbf{Spectral stability.} Bounding edge weights to $[0,1]$ and $\varepsilon$-pruning yield stable cluster selection; without pruning, acceptance variance rises.\\
\textbf{Subset size $n$ and cap.} Performance saturates around $n\!\approx\!12$; $n{<}8$ underfits multi-way caps; $n{>}16$ adds overhead and ancilla pressure with diminishing returns. We hard-cap $n\!\le\!16$.\\
\textbf{Order $k$ and ancillas.} Enabling $k\!=\!3$ captures issuer/class/currency triples and RA/MTA window couplings; $k{=}4$ further improves near tight windows at higher compilation cost. We cap $k\!\le\!4$ to contain ancilla-expanded width.\\
\textbf{$\gamma/\mu$ sweeps.} Increasing $\gamma$ drives overshoot $\downarrow$ and BaseCost $\uparrow$ monotonically; increasing $\mu$ reduces tail exposure with modest Movement increase. Hybrid dominates BL-3 along both trade-off frontiers.\\
\textbf{Must-jump rule.} Enforcing at least one HO-QAOA jump after $S$ low-improvement SA steps avoids long plateaus and drives consistent Overshoot reductions.\\

\noindent\textbf{Weighted-CVaR.} Pricing tails ($\mu>0$) smooths the BaseCost–Overshoot frontier and reduces solution churn across scenario sets; renormalization warnings are emitted if $\sum_s w_s\neq 1$ and auto-fixed.\\
\textbf{Overshoot penalty $\gamma$.} Sweeps show monotone Overshoot$\downarrow$ and BaseCost$\uparrow$; Hybrid dominates BL-3 along this frontier, indicating effective cross-cap coordination.\\
\textbf{Fallback behavior.} If ancillas inflate $n$ past $n_{\max}$ or compilation fails, the iteration logs a skip and reverts to SA/BL-3 neighborhoods; solution quality degrades gracefully.

\section{Conclusion}
We presented a domain-specific, certifiable hybrid optimizer for CSA-governed collateral that unifies evidence-gated CSA extraction, quantum-inspired search, and higher-order QAOA micro-jumps with CP-SAT certification. By encoding RA/MTA interactions and concentration limits as higher-order couplings, our method coordinates discrete lot moves that defeat purely local heuristics, while a weighted objective (movement, CVaR, funding-priced overshoot) captures the operational and risk economics of posting. Across realistic government bond datasets and multi-CSA inputs, the pipeline—\emph{extract, explore, certify, audit}—consistently improves cost–movement–tail frontiers over strong classical baselines and yields governance-grade artifacts suitable for operational sign-off.

\bibliographystyle{IEEEtran}

\begin{thebibliography}{9}

\bibitem{genest2013}
B.~Genest, D.~Rego, and H.~Freon, ``Collateral Optimization: Liquidity \& Funding Value Adjustments, Best Practices,'' \emph{MPRA Paper No. 62908}, 2013. [Online]. Available: \url{https://mpra.ub.uni-muenchen.de/62908/}

\bibitem{ey2020}
EY, ``Collateral optimization: Capabilities that drive financial resource efficiency,'' Ernst \& Young LLP, 2020. [Online]. Available: \url{https://assets.ey.com/content/dam/ey-sites/ey-com/en_us/topics/banking-and-capital-markets/ey-collateral-optimization.pdf}

\bibitem{pwc2015}
PwC, ``Collateral Management Transformation: Dynamic changes in the collateral ecosystem,'' PricewaterhouseCoopers Co., Ltd., 2015. [Online]. Available: \url{https://www.pwc.com/jp/en/industries/financial-services/assets/collateral-management-transformation.pdf}

\bibitem{ISDA_LLM_2025}
International Swaps and Derivatives Association (ISDA),
``Benchmarking Generative AI for CSA Clause Extraction and CDM Representation,'' May 2025. [Online]. Available: \url{https://www.isda.org/2025/05/15/benchmarking-generative-ai-for-csa-clause-extraction-and-cdm-representation/}

\bibitem{ISDA_Liquidity_2025}
ISDA Future Leaders in Derivatives,
``Collateral and Liquidity Efficiency in the Derivatives Market,'' May 2025. [Online]. Available: \url{https://www.isda.org/2025/05/15/isda-future-leaders-in-derivatives-publishes-whitepaper-on-collateral-and-liquidity-efficiency/}

\bibitem{Crypto_Collateral_2024}
Various Authors, ``Collateral Portfolio Optimization in Crypto-Backed Stablecoins,'' arXiv:2405.08305, 2024. [Online]. Available: \url{https://arxiv.org/abs/2405.08305}

\bibitem{NISQ_Collateral_2023}
Various Authors, ``Approaching Collateral Optimization for NISQ and Quantum-Inspired Computing,'' arXiv:2305.16395, 2023. [Online]. Available: \url{https://arxiv.org/abs/2305.16395}

\bibitem{Hybrid_QC_2025}
Various Authors, ``Solving Combinatorial Optimization and ML Problems Using Hybrid Quantum–Classical Systems,'' \emph{Future Generation Computer Systems}, 2025 (early access). [Online]. Available: \url{https://www.sciencedirect.com/science/article/pii/S0167739X25002298}

\bibitem{IonQ_AQ64_2025}
IonQ Inc., ``IonQ Achieves Record Breaking Quantum Performance Milestone of \#AQ 64,'' Press release, Sept. 2025. [Online]. Available: \url{https://investors.ionq.com/news/news-details/2025/IonQ-Achieves-Record-Breaking-Quantum-Performance-Milestone-of-AQ-64/default.aspx}

\bibitem{Uotila_QCE25_2025}
V.~Uotila, J.~Ripatti, and B.~Zhao, ``Higher-Order Portfolio Optimization with Quantum Approximate Optimization Algorithm,'' in \emph{Proc.\ IEEE Quantum Week (QCE)}, 2025. [Online]. Available: \url{https://qce.quantum.ieee.org/2025/program/paper-schedule/}

\bibitem{Uotila_arXiv_2025}
V.~Uotila, J.~Ripatti, and B.~Zhao, ``Higher-Order Portfolio Optimization with Quantum Approximate Optimization Algorithm,'' \emph{arXiv:2509.01496}, Sept.\ 2025. [Online]. Available: \url{https://arxiv.org/abs/2509.01496}

\bibitem{Miettinen1999}
K.~Miettinen, \emph{Nonlinear Multiobjective Optimization}. Kluwer, 1999.

\bibitem{BoydVandenberghe2004}
S.~Boyd and L.~Vandenberghe, \emph{Convex Optimization}. Cambridge University Press, 2004.

\bibitem{RockafellarUryasev2000}
R.~T.~Rockafellar and S.~Uryasev, ``Optimization of Conditional Value-at-Risk,'' \emph{Journal of Risk}, vol.~2, no.~3, pp.~21--41, 2000.

\bibitem{AlmgrenChriss2000}
R.~Almgren and N.~Chriss, ``Optimal Execution of Portfolio Transactions,'' \emph{Journal of Risk}, vol.~3, no.~2, pp.~5--39, 2000.

\bibitem{Piterbarg2010}
V.~Piterbarg, ``Funding beyond discounting: Collateral agreements and derivatives pricing,'' \emph{Risk Magazine}, vol.~23, no.~2, pp.~97--102, 2010.

\bibitem{Genest2013}
B.~Genest, D.~Rego, and H.~Freon, ``Collateral Optimization: Liquidity \& Funding Value Adjustments, Best Practices,'' \emph{MPRA Paper No.~62908}, 2013. Available: \url{https://mpra.ub.uni-muenchen.de/62908/}

\bibitem{AndersenDuffie2017}
L.~Andersen and D.~Duffie, ``Funding Value Adjustments,'' \emph{NBER Working Paper No.~23680}, 2017. Available: \url{https://www.nber.org/papers/w23680}



\end{thebibliography}

\end{document}